\documentclass[a4paper,12pt]{article}
\usepackage{amssymb}

\usepackage{hyperref}
\usepackage{subfigure}
\usepackage{ae}
\usepackage{mathrsfs}
\usepackage{graphicx}
\usepackage{bbm}
\usepackage{latexsym}
\usepackage{dsfont}
\usepackage{amsmath}
\usepackage[title,titletoc]{appendix}
\usepackage[super]{nth}
\usepackage{cancel}
\usepackage{booktabs}
\usepackage{multicol}
\usepackage{bbold}
\usepackage{amsfonts}
\usepackage{xcolor}
\usepackage[normalem]{ulem}
\newcommand{\mathsym}[1]{{}}

\newcommand{\qed}{\nobreak \ifvmode \relax \else \ifdim\lastskip<1.5em \hskip-\lastskip \hskip1.5em plus0em minus0.5em \fi \nobreak \vrule height0.75em width0.5em depth0.25em\fi}

\newcommand{\tr}{\mbox{tr}}

\def\app#1#2{  \mathrel{    \setbox0=\hbox{$#1\sim$}    \setbox2=\hbox{      \rlap{\hbox{$#1\propto$}}      \lower1.1\ht0\box0    }    \raise0.25\ht2\box2  }}

\begin{document}

	\begin{titlepage} 
		\begin{center} \hfill \\
		\hfill \\
			\textbf{\Large
		CP-odd and CP-even Weak-Basis Invariants in the Presence of Vector-Like Quarks}

			 \vskip 1cm \vskip 1cm Francisco Albergaria \footnote{francisco.albergaria@tecnico.ulisboa.pt}, G.C. Branco\footnote{gbranco@tecnico.ulisboa.pt},
José Filipe Bastos \footnote{jose.bastos@tecnico.ulisboa.pt} and
J.I. Silva-Marcos\footnote{juca@cftp.tecnico.ulisboa.pt}

\vskip 0.07in Centro de
F{\'\i}sica Te\'orica de Part{\'\i}culas, CFTP, \\ Departamento de
F\'{\i}sica,\\ {\it Instituto Superior T\'ecnico, Universidade de Lisboa, }
\\ {\it Avenida Rovisco Pais nr. 1, 1049-001 Lisboa, Portugal} \end{center}

		\begin{abstract} 
			We propose a minimal set of weak-basis invariants in an extension of the SM where one up-type isosinglet vector-like quark is introduced, which allows us to obtain all the physical content of the CKM matrix. We present CP-odd invariants of lower order in mass than the one in the SM, which may have important consequences for Baryogenesis. We study the extreme chiral limit, where the two lightest generations have vanishing mass, showing that in this extension, contrary to the SM, CP violation can be observed in collisions much above the electroweak scale.
\end{abstract}
	\end{titlepage}
	
\section{Introduction}
	
The addition of Vector Like Quarks (VLQ) to the spectrum of the Standard Model (SM) is one of the simplest extensions of the SM and VLQs may populate the desert between the SM and the Planck scale, without worsening the hierarchy problem. Among the best motivated VLQ models with either one down-type \cite{Belfatto2020} or one up-type \cite{Branco2021,Botella:2021uxz} isosinglet VLQ are those which attempt at solving the CKM unitarity problem \cite{Seng:2018yzq}-\cite{Aoki:2021kgd}. There are also models which feature both up and down-type isosinglet VLQs \cite{Belfatto2021, Crivellin2021} and models which feature VLQs in different $SU(2)$ representations \cite{Aguilar-Saavedra2013}.

The study of Weak-Basis Invariants (WBIs) has been carried out both in the quark sector \cite{Bernabeu1986}-\cite{Branco:1987mj} and the lepton sector \cite{Branco:1986} (see \cite{Wang:2021wdq} for a more recent analysis). 
In this paper, we study WBIs in an extension of the SM where an up-type $SU(2)$ singlet VLQ is introduced to the SM. In particular, we are interested in identifying a set of WBIs which enable one to reconstruct the full $4\times 3$ quark mixing matrix which arises in this extension of the SM. Let us recall that it has been shown \cite{Branco1988} that, in the SM with 3 generations and squared mass matrices $h_{i}=m_i m_i^\dagger, \ i=u,d$, for the up and down quarks, the WBIs $\text{tr}(h_u h_d)$, $\text{tr}(h^2_u h_d)$, $\text{tr}(h_u h^2_d)$ and $\text{tr}(h^2_u h^2_d)$, enable one to obtain the four independent moduli of the CKM matrix. With the knowledge of these moduli, one can reconstruct the full CKM, including
the strength of CP violation. There is only a two-fold ambiguity in the sign of CP violation which can be lifted by the evaluation of the CP-odd WBI $\text{tr}[h_u,h_d]^3$ \cite{Bernabeu1986}.  
   
Here, we analyse the analogous question in the context of an extension of the SM with an up-type VLQ. In this case there are nine independent moduli and we identify the nine WBIs which enable one to reconstruct the full quark mixing matrix. Special attention is given to the extreme chiral limit (ECL) where the first two generations may be considered massless. In this limit, there is no CP violation in the SM. We will show that in the VLQ extension there is CP violation, even in the ECL, and identify CP-odd WBI which do not vanish in this limit. 

% ---------------------------------

One of the reasons why in the SM it is not possible to generate a Baryon Asymmetry
of the Universe (BAU) consistent with observation, has to do with the fact that
CP-violation in the SM is too small. A rough estimate is
\begin{equation}\label{I_SM}
    \mathcal{I}_\text{CP} =\text{tr}[y_uy^\dagger_u, y_dy^\dagger_d]^3 \sim  \frac{\text{tr}[h_u, h_d]^3}{v^{12}} \sim 10^{-25},
\end{equation}
%
% I computed this using eq. 14.24 of the book by Gustavo, Lavoura and Silva
%
where $y_u$ and $y_d$ are the Yukawa-coupling matrices, which are related to the quark mass matrices as $y_u = \frac{1}{v} m_u$, $y_d = \frac{1}{v} m_d$ and $v$ is the electro-weak scale.
The SM CP-odd WBI, $\text{tr}[h_u, h_d]^3$ has a mass order of $M^{12}$. However, in our extension of the SM with the up-VLQ there is a CP-odd WBI of much lower order in mass, namely, of order $M^8$.
% Therefore one expects:
%
%                               I(CP)VLQ= ...
%
%
Therefore, one expects that the contribution to BAU in the framework of models with VLQs to be much larger than in the SM.

% -----------------------------

The paper is organised as follows: in section \ref{sec:WBI} we present a set of WBIs which allow us to obtain all the information contained in the CKM matrix. In section \ref{sec:CPV}, we study CP violation (in this extension of the SM) and show that we get a much bigger contribution to CP violation than in the SM due to lower order in mass CP-odd invariants. In section \ref{sec:ECL}, we study the extreme chiral limit, where at extremely high energies it is considered that $m_u = m_c = 0$ and $m_d = m_s = 0$, present a parametrization for the CKM matrix in this limit and show that we have CP violation, even at these energies. In the Appendices, we give a general setup of invariants for more than one VLQ and construct a simplified set of invariants completely characterizing all parameters for the case of one up-VLQ. Finally, in section \ref{sec:conclusion}, we present our conclusions.

	\section{Weak-Basis Invariants}\label{sec:WBI}
	
We consider a model with $3$ families of SM-like quarks and one additional up-type isosinglet VLQ (1-VLQ). 
In this model there are 16 physical parameters in the quark sector: $4$ related to the masses of the up-type quarks, $3$ related to the masses of the down-type quarks and $9$ related to the mixing.
Following \cite{Branco2021,Botella:2021uxz}, after spontaneous symmetry breaking, we write the quark-mass Lagrangian as
\begin{equation}
    \mathcal{L}_M = - \overline{u}_{Li} (m_u)_{i \alpha} u_{R \alpha} - \overline{U}_{L 1} (M_u)_{1 \alpha} u_{R \alpha} - \overline{d}_{L i} (m_d)_{ij} d_{R j},
\end{equation}
where the indices $i$ and $j$ run from $1$ to $3$ and the index $\alpha$ runs from $1$ to $4$, $u_{L i}$ denote the $3$ left-handed up-type SM quarks, $u_{R \alpha}$ are the $4$ right-handed up-type quarks (the right-handed up-type SM quarks and the right-handed VLQ are indistinguishable as they have the same quantum numbers), $U_{L 1}$ is the $1$ left-handed up-type VLQ, $d_{Li}$ are the $3$ left-handed down-type SM quarks and $d_{Rj}$ are the $3$ right-handed down-type SM quarks. The matrix $m_u$ is a $3 \times 4$ matrix, the matrix $M_u$ is a $1 \times 4$ matrix (\textit{i.e.}, it is a row vector) and the matrix $m_d$ is a $3 \times 3$ matrix. The $4 \times 4$ up-type quark mass matrix is given by
\begin{equation}
    \mathcal{M}_u = \begin{pmatrix}
m_u \\
M_u
\end{pmatrix}.
\end{equation}

\subsection{Weak-Basis tansformations}

In order to construct the WB invariants, we have to specify the WB tranformations. 
Under a general weak-basis transformation, the quark fields  transform as
\begin{equation}
\begin{array}{lll}
u_{L}\rightarrow V\ u_{L} &  & d_{L}\rightarrow V\ d_{L} \\ 
u_{R}\rightarrow \mathcal{W}_{u}\ u_{R} &  & d_{R}\rightarrow W_{d}\
d_{R} \\ 
U_{L}\rightarrow Z\ U_{L} &  & 
\end{array}
\label{transf}
\end{equation}
where the unitary $V$, $W_{d}$ are $3 \times 3$ matrices, while $\mathcal{W}_{u}$ is a $4 \times 4$ unitary matrix. Notice that $V$ appears simultaneously in the left-handed fields of the up and down sector, thus the charged current Lagrangian term is also left invariant.  For one up-VLQ, $Z$ is just some complex phase factor $e^{i \alpha}$, however here, we maintain a matrix-form formulation for $Z$, which can be easily generalized to a matrix in the case of more than one VLQ's. 

These transformation properties for the quark fields imply that the mass matrices of the quarks transform under a weak-basis transformation as
\begin{equation}
\begin{array}{lllll}
m_{u}\rightarrow V^{\dagger }m_{u}\mathcal{W}_{u} & , & 
M_{u}\rightarrow Z^{\dagger }M_{u}\mathcal{W}_{u} & , & m_{d}\rightarrow
V^{\dagger }m_{d}W_{d}.
\end{array}
\label{transf_2}
\end{equation}
Hence, we have
\begin{equation}
\begin{array}{lll}
m_{u}m_{u}^{\dagger }\rightarrow V^{\dagger }m_{u}m_{u}^{\dagger }V
&  & m_{d}m_{d}^{\dagger }\rightarrow V^{\dagger }m_{d}m_{d}^{\dagger
}V \\ 
&  &  \\ 
m_{u}M_{u}^{\dagger }M_{u}m_{u}^{\dagger }\rightarrow V^{\dagger
}m_{u}M_{u}^{\dagger }M_{u}m_{u}^{\dagger }V &  & m_{d}^{\dagger
}m_{d}\rightarrow W_{d}^{\dagger }m_{d}^{\dagger }m_{d}W_{d} \\ 
&  &  \\ 
M_{u}M_{u}^{\dagger }\rightarrow Z^{\dagger }M_{u}M_{u}^{\dagger }Z &  &  \\ 
&  &  \\ 
m_{u}^{\dagger }m_{u}\rightarrow \mathcal{W}^{\dagger }_{u}m_{u}^{\dagger
}m_{u}\mathcal{W}_{u}, &  & M_{u}^{\dagger }M_{u}\rightarrow \mathcal{W}%
_{u}^{\dagger }M_{u}^{\dagger }M_{u}\mathcal{W}_{u}
\end{array}
\label{eq:mmtransform}
\end{equation}

Taking traces of appropriate combinations of the matrices in Eq. \eqref{eq:mmtransform}, one obtains different weak-basis invariant quantities. We must now find a minimal (and comprehensive) set of these quantities, such these can be easily related to the elements of the CKM matrix and to the quark masses, and thus obtain adequate information which relates both.

\subsection{Constructing the Invariants}\label{sec:constr_invs}
% In order to relate the weak-basis invariants with the elements of the CKM matrix, it is useful to consider certain convenient weak basis (WB). In the following, we choose a WB in which the down quark mass matrix is \text{diag}onal. When \text{diag}onalizing the up-quark mass matrix, to obtain the quark masses, one also obtains the charged current and, because of the VLQ, also neutral current coupling elements. 

The up quark mass matrix is \text{diag}onalized by a bi-unitary transformation of the fields such that
\begin{equation}
\begin{array}{lll}
\left( 
\begin{array}{l}
u \\ 
U
\end{array}
\right) _{L}\rightarrow \mathcal{V\ }u'_{L} & , & u_{R\mathcal{\ }}\rightarrow \mathcal{W\ }u_{R\mathcal{\ }%
}^{\prime }
\end{array}
\end{equation}
where $\mathcal{V},\ \mathcal{W}$ are unitary $4 \times 4$ matrices and
\begin{equation}
   \mathcal{V}^\dagger  \mathcal{M}_u \mathcal{W}=\mathcal{D}_u.
   \label{Du}
\end{equation}
Defining for the up-quarks, $\mathcal{H}_u = \mathcal{M}_u \mathcal{M}_u^\dagger$, we have four invariants related to the masses
\begin{equation}
\begin{split}
\text{tr}(\mathcal{H}_{u})=& \ m_{u}^{2}+m_{c}^{2}+m_{t}^{2}+m_{T}^{2}, \\ 
\\ 
\chi _{1}(\mathcal{H}_{u})\equiv & \ \frac{1}{2}\frac{\partial ^{2}}{\partial
x^{2}}\det (\mathcal{H}_{u}-x\mathbb{1}_{4\times 4})|_{x=0} \\ 
=& \ m_{u}^{2}m_{c}^{2}+m_{u}^{2}m_{t}^{2}+m_{u}^{2}m_{T}^{2}+m_{c}^{2}m_{t}^{2}+m_{c}^{2}m_{T}^{2}+m_{t}^{2}m_{T}^{2},
\\ 
\\ 
\chi _{2}(\mathcal{H}_{u})\equiv & \ - \frac{\partial }{\partial x}\det (\mathcal{%
H}_{u}-x\mathbb{1}_{4\times 4})|_{x=0} \\ 
=& \ m_{u}^{2}m_{c}^{2}m_{t}^{2}+m_{u}^{2}m_{c}^{2}m_{T}^{2}+m_{u}^{2}m_{t}^{2}m_{T}^{2}+m_{c}^{2}m_{t}^{2}m_{T}^{2},
\\ 
\\ 
\det (\mathcal{H}_{u})=& \ m_{u}^{2}m_{c}^{2}m_{t}^{2}m_{T}^{2},
\end{split}
\label{eq:WBImasses}
\end{equation}
where $m_u$, $m_c$ and $m_t$ are the masses of the up-type SM quark and $m_T$ is the (heavy) mass of the up-type VLQ. A similar set of three invariants defines the down quark masses, where instead of $\chi_1$ and $\chi_2$ we just have
\begin{equation}
    \chi(h_d)\equiv -\frac{\partial}{\partial x}\text{det}\left(h_d-x\mathbb{1}_{3\times 3}\right)|_{x=0}=m_{d}^{2}m_{s}^{2}+m_{d}^{2}m_{b}^{2}+m_{s}^{2}m_{b}^{2}.
\end{equation}
where $h_d$ is the equivalent $3\times 3$ squared down-quark matrix.

We write the unitary matrix which \text{diag}onalizes the up-quark matrix (on the left) $\mathcal{V}$ 
as
\begin{equation}
    \mathcal{V} = \begin{pmatrix}
A \\
B
\end{pmatrix},
\label{AB}
\end{equation}
where $A$ and $B$ are $3 \times 4$ and $1 \times 4$ matrices, respectively. With these, we obtain (in the WB where the down-quark matrix is \text{diag}onal) the charged current Lagrangian 
\begin{equation} \label{eq:chargedcurrent}
    \mathcal{L}_W = \frac{g}{\sqrt{2}} W_\mu^+ \overline{u^\prime}_{L}\  A^\dagger\  \gamma^\mu d^\prime_L + \mathrm{h.c.}
\end{equation}
Thus we find that (in this model), the CKM matrix is given by the $4 \times 3$ non-unitary matrix
\begin{equation}
V_\text{CKM}=A^\dagger.
\end{equation} 
Furthermore, we obtain for the neutral current Lagrangian 
\begin{equation} \label{eq:neutralcurrent}
    \mathcal{L}_Z = \frac{g}{c_W} Z_\mu \left[ \frac{1}{2} \left(\overline{u^\prime}_{L}\  A^\dagger A\  \gamma^\mu u^\prime_L - \overline{d^\prime}_{L} \gamma^\mu d^\prime_L \right) - s_W^2 \left(\frac{2}{3} \overline{u^\prime} \gamma^\mu u^\prime - \frac{1}{3} \overline{d^\prime} \gamma^\mu d^\prime \right) \right],
\end{equation}
which means that, in the charge $2/3$ up-quark sector, we will have Flavour-Changing Neutral Currents (FCNCs) at tree-level. We define the FCNC-matrix 
\begin{equation}F^u\equiv A^\dagger A= V_\text{CKM}V^\dagger_\text{CKM},
\end{equation}
which, as known, in the SM is equal to the identity. However, here and in general for VLQ-models, it is not a diagonal matrix, thus leading to the FCNCs.

Due to the unitarity of $\mathcal{V}$ we have
\begin{equation}
\begin{array}{lll}
AA^{\dagger }=\mathbb{1}_{3\times 3}, &  & A^{\dagger }A+B^{\dagger }B=%
\mathbb{1}_{4\times 4}, \\ 
&  &  \\ 
BB^{\dagger }=1, &  & AB^{\dagger }=0\Leftrightarrow BA^{\dagger }=0.
\end{array}
\label{eq:AAdagger}
\end{equation}
Note that, although the CKM matrix is no longer unitary (in fact, it is not even a square matrix), we do still have $V_\text{CKM}^\dagger V_\text{CKM} = \mathbb{1}_{3 \times 3}$.

From Eqs. (\ref{Du}, \ref{AB}), the \text{diag}onalized up-type quark mass matrix $\mathcal{D}_u$ is given by
\begin{equation} \label{eq:Du}
    \mathcal{D}_u = A^\dagger m_u \mathcal{W} + B^\dagger M_u \mathcal{W}.
\end{equation}
By multiplying Eq. \eqref{eq:Du} on the left by $A$ or $B$ and on the right by $\mathcal{W}^\dagger$, and using the identities from Eq. \eqref{eq:AAdagger}, we obtain
\begin{equation}
 m_u = A\  \mathcal{D}_u \mathcal{W}^\dagger \ ,\ \ M_u = B\  \mathcal{D}_u \mathcal{W}^\dagger
 \label{muMu}
\end{equation}

\subsubsection*{A Set of Hermitian Matrices relating Physical Parameters}

Now, consider the following Hermitian matrices
\begin{equation}%\label{eq:hermitian}
    \begin{array}{c}\label{eq:herm_matr}
{H}^{(r)}_u\equiv  m_u (m_u^\dagger m_u  +  M_u^\dagger M_u)^{r-1} m_u^\dagger,\\
    h_d\equiv m_dm^\dagger_d
    \end{array}
\end{equation}
All these Hermitian matrices transform under weak-basis transformations (WBT) in (\ref{transf}) as
\begin{equation}
    h \rightarrow V^\dagger\   h\  V.
    \label{uuu}
\end{equation}
From Eqs. (\ref{eq:AAdagger}, \ref{muMu}), in the WB where the down sector mass matrix is \text{diag}onal, these verify
\begin{equation}
    \begin{array}{c}
    {H}^{(r)}_u= A (\mathcal{D}^{2}_u)^r A^\dagger,\\
    h_d=  \text{\text{diag}}(m^2_d,m^2_s,m^2_b)
    \end{array}
\end{equation}
and, with $A^\dagger=V_\text{CKM}$, we find the following relations between the weak-basis invariants and the CKM matrix moduli and the quark masses
\begin{equation} \label{eq:WBI}
    \tr \left({H}^{(r)}_u h^s_d \right) = \sum_{\alpha=1}^{4} \sum_{i = 1}^3 (m_{u_\alpha}^{2})^r (m_{d_i}^{2})^s |V_{\alpha i}|^2, 
\end{equation}
where $s$ and $r$ are integers. 

Henceforth we will use the simplified notation\footnote{Note that ${H}^{(r)}_u$ corresponds to the upper-left $3\times 3$ block of $\mathcal{H}^r_u$. The analogous quantity in the SM would be $h^r_u=(m_um^\dagger_u)^r$, but for these type of extensions, for $r\neq 1$, these are strictly different, i.e. $h^r_u\neq {H}^{(r)}_u$, even though both transform in the same way under WBTs.} $h_u\equiv {H}^{(1)}_u=m_u m^\dagger_u$.

Finally, we compute $9$ invariants obtained from Eq. \eqref{eq:WBI}. To do this, we choose the pair $(r,s)$ to be equal to $(1,0)$, $(1,1)$, $(1,2)$, $(1,3)$, $(2,0)$, $(2,1)$, $(2,2)$, $(3,0)$ and $(3,1)$. Knowing the masses of the quarks through the weak-basis invariants in Eq. \eqref{eq:WBImasses}, we obtain $9$ independent equations of order equal or lower than $M^8$ for the $9$ independent moduli of the $4 \times 3$ CKM matrix elements. From the knowledge of the $9$ independent moduli, it is possible to know all the information contained in the CKM matrix
up to a discrete ambiguity\footnote{This discrete ambiguity is similar to what happens in the SM, where, when computing the four WBIs which determine the four independent moduli of the CKM matrix, we still have to fix the ambiguity of the sign of CP violation.},
as it was done in \cite{Lavoura1989}. 

Thus, up to a discrete ambiguity, we obtain all the information contained in the CKM matrix by computing these $9$ invariants.

As an example, in Appendix \ref{apa}, we have explicitly constructed a comprehensive set of simplified invariants (which can be easily verified) and which completely characterizes all parameters of this extension of the SM with one up-VLQ. For a general setup concerning invariants for more than one VLQ, we refer to  Appendix \ref{WBIgen}.

% In a model with more VLQs, we could still use these weak-basis invariants to obtain equation for the masses of the quarks but we would need more invariants which we could relate to the masses of the quarks.

\section{CP Violation and Consequences for Baryogenesis}\label{sec:CPV}

The argument we used in the last section to show that one can determine all parameters in an extension with a VLQ isosinglet in terms of WBIs was solely dependent on real quantities.
However, the existence of CP violation (CPV) hinges on the existence of CP-odd WBIs. 

In the SM the simplest (\textit{i.e.} lowest order) quantity with such a property is \cite{Bernabeu1986}
\begin{equation}
    \begin{split}
    \text{tr}\left[h_u, h_d\right]^3 = 6i &(m^2_t-m^2_c) (m^2_t-m^2_u) (m^2_c-m^2_u)\\
    & {\times} (m^2_b-m^2_s) (m^2_b-m^2_d) (m^2_s-m^2_d) I_\text{CP},
    \end{split}
    \label{CP_SM}
\end{equation}
where
\begin{equation}
    I_\text{CP}\equiv \text{Im}\left(V_{\alpha i}V_{\beta j}V^*_{\alpha j}V^*_{\beta i}\right)
\end{equation}
and $V_{\alpha i}V_{\beta j}V^*_{\alpha j}V^*_{\beta i}$ correspond to any invariant quartet ($i\neq j$, $\alpha\neq \beta$), which in the SM all have the same imaginary part up to a sign. This invariant \cite{Branco:1987mj,Jarlskog1988} was suggested by  Jarlskog and Stora and independently by Branco and Lavoura, with both papers appearing in the same volume of \textit{Phys. Lett. B}.

When VLQs are introduced to the theory, in general more physical phases are present. For instance, in the case of the SM extended with an up-type VLQ isosinglet ($N=1$) there exist three physical phases and therefore extra sources of CP violation. This then results in a larger set of necessary and sufficient conditions for CP invariance. 

From \cite{delAguila:1997vn}, a minimal set of seven CP-odd WBI is:
\begin{equation}\label{CP-odd}
\begin{split}
     \text{tr}\left(\left[h_u, h^s_d\right]H^{(2)}_u\right)&=2i \hspace{0.5mm} m^{2s}_{d_i}m^4_{u_\alpha}m^2_{u_\beta}\text{Im}\left(F^u_{\alpha \beta}V^*_{\alpha i}V_{\beta i}\right), \\
    \text{tr}\left(\left[h^2_u, h^s_d\right]H^{(2)}_u\right)&=2i \hspace{0.5mm} m^{2s}_{d_i}m^4_{u_\alpha}m^2_{u_\beta}m^2_{u_\rho}\text{Im}\left(F^u_{\alpha\rho}F^u_{\rho\beta}V^*_{\alpha i}V_{\beta i}\right),\\
    \text{tr}\left(\{h_u, h_d\}[h^2_d,H^{(2)}_u]\right)&=2i \hspace{0.5mm} m^4_{d_i}m^2_{u_\alpha}m^4_{u_\beta} \text{Im}\left[V_{\alpha i}V^*_{\beta i}\left(m^2_{d_j}V^*_{\alpha j}V_{\beta j}-m^2_{d_i}F^u_{\beta\alpha}\right)\right],\\
    \text{tr}\left(\{H^{(2)}_u, h^2_d\}[h_d,H^{(2)}_u]\right)&=2i \hspace{0.5mm} m^4_{d_i}m^2_{u_\alpha}m^2_{u_\beta}m^4_{u_\rho} \text{Im}\left [V_{\alpha i}V^*_{\rho i}F^u_{\beta\alpha}\left(m^2_{d_j}V^*_{\beta j}V_{\rho j}-m^2_{d_i}F^u_{\rho\beta}\right)\right],\\
    \text{tr}\left[h_u, h_d\right]^3&=6i \hspace{0.5mm} m^{2}_{d_i}m^{4}_{d_j}m^2_{u_\alpha}m^2_{u_\beta}m^2_{u_\rho}\text{Im}\left(V_{\alpha i}V_{\beta j}V^*_{\alpha j}V^*_{\rho i}F^u_{\rho\beta}\right),
\end{split}
\end{equation}
for $s=1,2$ and with an implicit sum over the quark indices (with $\alpha,\beta,\rho=1,2,3,4$ and $i,j=1,2,3$). The vanishing of this set of CP-odd invariants is a necessary and sufficient condition for CP invariance.

The fact that CP violation now depends on a variety of invariants instead of a single one could mean that in the presence of a VLQ there exists an enhancement of CP violation, since even if the standard CP-odd invariant vanished, \text{i.e.} $\text{tr}[h_u,h_d]^3=0$, there could exist CP violation arising from other sources. Not only that, but in principle these new sources could lead to a much larger CP violation than the one predicted in the SM. This is because now there are several CP-odd WBIs of dimension lower than the SM invariant in Eq. (\ref{CP_SM}), of dimension $M^{12}$. In fact, the invariant of lowest dimension is
$\text{tr}\left(\left[h_u, h_d\right]H^{(2)}_u\right)$, of dimension $M^8$ which is necessarily\footnote{In this extension, when building WBIs using hermitian matrices, as for instance $h_u$ and $h_d$, as building blocks, all resulting invariants have even mass dimension $M^{2n}$. Now, any invariant of dimension $M^2$ is simply the trace of an hermitian matrix, therefore real and providing no information regarding CPV. For dimension $M^4$ invariants the problem persists, because the product of two hermitian matrices still has real trace. Similarly, dimension $M^6$ invariants can, in this case, also only be written as products of two hermitian matrices, e.g. $\tr(h^2_d h_u)$ or $\tr(h_d H^{(2)}_u)$, leaving us with $M^8$ as the lowest possible dimension for CP-odd invariants.} the lowest order for CP-odd invariants in these type of theory.

Similarly to the SM quantity in Eq. (\ref{I_SM}), we expect that for a model with a VLQ, the size of CP violation is dominated by a dimensionless quantity such as
\begin{equation}
    \mathcal{I}_\text{CP}=\frac{\text{tr}\left(\left[h_u, h_d\right]H^{(2)}_u\right)}{V^8},
\label{I_CP}
\end{equation}
In the evaluation of this WBI and since we are considering a model beyond the SM, we also could expect to have some energy scale $V$ different from $v$ of the SM. Reasonably, one may choose this scale $V\sim m_T$, the mass of the VLQ, or attending to the constituent matrix elements of $\text{tr}\left(\left[h_u, h_d\right]H^{(2)}_u\right)$ in Eq. (\ref{eq:herm_matr}), one may also consider $V^8\sim v^6 m_T^2$.
In each case, our new quantity in Eq. (\ref{I_CP}) has the potential to be much larger than the SM quantity. 
This could provide a significant enhancement of CP violation in the quark sector, which in turn could be useful in explaining the observed Baryon Asymmetry of the Universe\footnote{Note that this enhancement of CP violation, possible in this type of extension, only fulfills one of the necessary conditions for Baryogensesis. There is also the requirement of a strong first-order phase transition, however this topic is outside the scope of this work centered around the study of WBIs.}. A similar argument could even be extended to the next-lowest dimension invariants, those of dimension $M^{10}$, such as $\text{tr}\left(\left[h_u, h^2_d\right]H^{(2)}_u\right)$ and $\text{tr}\left(\left[h^2_u, h_d\right]H^{(2)}_u\right)$.

\subsection{The size of CPV with one VLQ: a Numerical Example}

Next, we provide a concrete numerical example of the size of CPV with one VLQ. We consider case II of Appendix B in \cite{Botella:2021uxz}. This example is relevant because it is associated to an extension of the SM with an up-type VLQ isosinglet, which addresses the CKM unitarity problem and accommodates the various tests posed by the electroweak measurements, particularly newly derived stringent constraints set by the CPV parameter $\varepsilon_K$.   
We recall the up-quark mass matrix of the case, (given in the WB where the down-sector mass matrix is diagonal), 
\begin{equation}
		\mathcal{M}_{u}= \left(
        \begin{array}{cccc}
         0 & 0 & 0 & 65.033 \\
         0 & 0 & 7.12124 & 15.8436 e^{1.92462 i} \\
         0 & 19.3672 & 172.73 & 4.21828 e^{-1.56762 i} \\
         0.0397187 & 1.63403 & 32.7938 e^{-1.51551 i} & 1475.32 \\
        \end{array}
        \right) \text{GeV},
\end{equation}
at the $M_Z$ scale, with leads to the quark masses (in GeV)
\begin{equation}
		\begin{array}{cccc}
			m_{u}=0.0018, & m_{c}=0.77, &
			m_{t}=174 , & m_T=1477.
		\end{array}
		\label{spect}
	\end{equation}
and the (full) quark mixing matrix
\begin{equation}
	\small
		|\mathcal{V}^{\dagger }|\simeq\left(
        \begin{array}{cccc}
         0.973609 & 0.223644 & 0.00382359 & 0.0453188 \\
         0.223785 & 0.973837 & 0.0395133 & 0.000740405 \\
         0.00824668 & 0.0388324 & 0.999212 & 0.000234355 \\
         0.0440136 & 0.010821 & 0.000312045 & 0.998972 \\
        \end{array}
        \right),
\end{equation}
with $I_\text{CP}\simeq 3.070\times 10^{-5}$.

Using the down-quark (in GeV)
\begin{equation}
    m_d=0.003, \quad \quad m_s=0.066, \quad \quad m_b=2.90,
\end{equation}
we obtain 
\begin{equation}
    \left|\mathcal{I}_\text{CP}\right|=\left|\frac{\text{tr}\left(\left[h_u, h_d\right]H^{(2)}_u\right)}{m^2_Tv^6}\right| \simeq 2.02\times 10^{-10},
\end{equation}
which is to be compared with the SM estimate in Eq.(\ref{I_SM}).

Furthermore, if we consider one other of the invariants of dimension $M^{10}$ we find

\begin{equation}
    \begin{split}
    \left|\mathcal{I}^{'}_\text{CP}\right|\equiv & \ \left|\frac{\text{tr}\left(\left[h^2_u, h_d\right]H^{(2)}_u\right)}{m^2_Tv^{8}}\right| \simeq 1.16\times 10^{-10},\\
    \end{split}
\end{equation}
which may also be relevant. The same is not true for the remaining invariants in Eq. (\ref{CP-odd}).

\section{The Extreme Chiral Limit} \label{sec:ECL}

An important and interesting result with regard to CP violation and VLQ's comes from the study of physical processes at very high energies. At these energies, the masses of the first two light quark generations are somewhat irrelevant and we may consider them insignificant or even to vanish. We refer to this situation as the extreme chiral limit (ECL). In this limit, there is only one independent CP-odd WBI \cite{delAguila:1997vn}. The vanishing of this invariant implies CP invariance.

Let us consider in this limit the following WB, where without loss of generality, the quark mass matrices assume the following form for
\begin{equation}
    m_d= \text{\text{diag}}(0,0,m_b), \quad \quad \mathcal{M}_u'=\mathcal{V}\cdot \text{\text{diag}}(0,0,m_t,m_T),
    \label{ECL_mass_matrices}
\end{equation}
$\mathcal{V}$ is a $4\times 4$ unitary matrix and the CKM mixing is the $4\times 3$ non-unitary left part of $\mathcal{V}^\dagger$.

In this limit and specific WB, one has from Eq. (\ref{ECL_mass_matrices}), that
\begin{equation}
    \mathcal{M}_u'=\begin{pmatrix}
    0 & 0 & m^c_{13} & m^c_{14}\\
    0 & 0 & m^c_{23} & m^c_{24}\\
    0 & 0 & m^c_{33} & m^c_{34}\\
    0 & 0 & m^c_{43} & m^c_{44}
    \end{pmatrix},
\end{equation}
where, in general, the $m_c$'s are complex matrix elements. 

Furthermore, one can always make a unitary transformation of the right-handed quark fields, e.g. in the $(3,4)-$plane, $\mathcal{W}^R_{34}$, since any of these unitary transformation is unphysical.
Additionally, because of the vanishing of quark masses for the first two generations, one may also redefine the left-handed quark fields through an arbitrary unitary transformation in the $(1,2)-$plane, $\mathcal{U}^L_{12}$.
Apart from these transformations, and given the fact that the down-sector is already diagonal, one has also the freedom of rephasing, not only all the RH quark fields, but also the LH quark fields.

Under all these transformations $\mathcal{M}_u'$ becomes
\begin{equation}
    \mathcal{M}_u'\rightarrow\mathcal{M}_u=(\mathcal{K}_L \mathcal{U}^L_{12}) \cdot\mathcal{M}_u'\cdot (\mathcal{W}^R_{34}\mathcal{K}_R),
\end{equation}
where the $\mathcal{K}_{L,R}$ are diagonal matrices of phases accounting for the rephasings.

The freedom in choosing $\mathcal{W}^R_{34}$ and $\mathcal{U}^L_{12}$ allows one to impose $m_{43}=m_{13}=0$, meaning that, in this limit, there are only 6 independent mass matrix entries, and one may choose the (only) complex phase of $\mathcal{M}_u$ such that
\begin{equation}
\mathcal{M}_{u}=\left( 
\begin{array}{llll}
0 & 0 & 0 & m_{14} \\ 
0 & 0 & m_{23} & m_{24}e^{i\varphi } \\ 
0 & 0 & m_{33} & m_{34} \\ 
0 & 0 & 0 & m_{44}
\end{array}
\right) ,  \label{Muprime}
\end{equation}
with all the $m_{ij}$ real. 

With this minimal form in mind, it is straightforward to check that (after factoring out all unnecessary and un-physical phase factors), one obtains for unitary matrix $\mathcal{V}^\dagger$ which diagonalizes $\mathcal{M}_u$ as in Eq. \eqref{Du},
\begin{equation}\label{V_ECL}
    \begin{split}
    \mathcal{V}^\dagger=& \ \mathcal{O}_{34}\mathcal{K}_{\delta_\text{CL}} \mathcal{O}_{24}\mathcal{O}_{14} \mathcal{O}_{23}=\\
    = & \ \begin{pmatrix}
    1 & 0 & 0 & 0\\
    0 & 1 & 0 & 0\\
    0 & 0 & c_{34} & s_{34}\\
    0 & 0 & -s_{34} & c_{34}\\
    \end{pmatrix}
    \begin{pmatrix}
    1 & 0 & 0 & 0\\
    0 & 1 & 0 & 0\\
    0 & 0 & 1 & 0\\
    0 & 0 & 0 & e^{i\delta_\text{CL}}\\
    \end{pmatrix}
    \begin{pmatrix}
    1 & 0 & 0 & 0\\
    0 & c_{24} & 0 & s_{24}\\
    0 & 0 & 1 & 0\\
    0 & -s_{24} & 0 & c_{24}\\
    \end{pmatrix}\\
    \cdot & \ \begin{pmatrix}
    c_{14} & 0 & 0 & s_{14}\\
    0 & 1 & 0 & 0\\
    0 & 0 & 1 & 0\\
    -s_{14} & 0 & 0 & c_{14}\\
    \end{pmatrix}\begin{pmatrix}
    1 & 0 & 0 & 0\\
    0 & 1 & 0 & 0\\
    0 & 0 & c_{23} & s_{23}\\
    0 & 0 & -s_{23} & c_{23}\\
    \end{pmatrix},
    \end{split}
\end{equation}
where $\delta_\text{CL}$ is a function of the phase $\varphi$ and the real parameters in Eq.(\ref{Muprime}).

Finally, one finds the CKM matrix for the ECL, corresponding to the first three columns of $\mathcal{V}^\dagger$, i.e.
\begin{equation}\label{eq:ECL_param}
    V_\text{CKM}=\begin{pmatrix}c_{14} & 0 & 0 \\
 -s_{14} s_{24} & c_{23} c_{24} & s_{23} c_{24} \\
 - s_{14} c_{24} s_{34} e^{i \delta_\text{CL} } & -s_{23}c_{34}- c_{23} s_{24} s_{34} e^{i \delta_\text{CL} } & c_{23} c_{34}- s_{23} s_{24} s_{34} e^{i \delta_\text{CL} }\\
 - s_{14} c_{24} c_{34} e^{i \delta_\text{CL} } & s_{23} s_{34}- c_{23} s_{24} c_{34} e^{i \delta_\text{CL} } & -c_{23} s_{34}- s_{23} s_{24} c_{34} e^{i \delta_\text{CL} } \\
   \end{pmatrix},
\end{equation}
with $c_{ij}\equiv \cos(\theta_{ij})$, $s_{ij}\equiv\sin(\theta_{ij}) \in [0,\pi/2]$ and $\delta_\text{CL}\in[0,2\pi]$. 

In the ECL, there exists only one independent CP-odd invariant, as already was implied in \cite{delAguila:1997vn}, and we may now write all of the invariants in Eq.(\ref{CP-odd}) in terms of the invariant of lowest dimension,
\begin{equation}
   \text{tr}\left(\left[h_u, h_d\right]H^{(2)}_u\right)= 2i \hspace{0.5mm} m^2_b m^2_t m^2_T(m^2_T-m^2_t)I_\text{CL},
\end{equation}
where the ECL rephasing invariant yields
\begin{equation}
I_\text{CL}=\text{Im}\left(F^u_{34}V^*_{33}V_{43}\right)=c_{24}c^2_{14}c^2_{24}c_{34}s_{23}s_{24}s_{34}\sin\delta_\text{CL}.
\end{equation}

We point out that even at very high energies there might exist new important sources of CPV. This contrasts with the SM, where there is no CP violation in the ECL, as can be straightforward verified with the SM-CP-odd invariant in Eq.(\ref{CP_SM}), proportional to $m^2_c-m^2_u$ and $m^2_s-m^2_d$, and which vanishes in this limit. 

A rough estimate of the size of CPV at very high energies leads to the upper-bound
\begin{equation}
    \mathcal{I}_\text{CL}=\left[\frac{\text{tr}\left(\left[h_u, h_d\right]H^{(2)}_u\right)}{m^2_Tv^6}\right]_\text{ECL}\lesssim \frac{m^2_b m^2_T}{m^4_t}\left|s_{23}s_{24}s_{34}\right|.
\end{equation}
for small mixing angles.

Once more, these results demonstrate that the existence of VLQs could have very important implications for CPV.
Nonetheless, the experimental measurement of CPV at extremely high energies is a notably challenging task \cite{Bernabeu1989,Hou1986}.

	\section{Conclusions} \label{sec:conclusion}
	
We have studied CP-even and CP-odd WB invariants in the context of extensions of the SM where vector-like quarks are added to the spectrum of the SM. 

WBIs are very 
useful for studying some of the physical implications of VLQ's. For example, it is well known that in high energy collisions one cannot distinguish a $d$-quark jet from 
a $s$-quark jet. Therefore, at high energy collisions (\textit{i.e.}, energies much higher than $v$) in the context of the SM, CP violation effects vanish. However, by constructing CP-odd invariants which, in the presence of VLQ do not vanish, one can show that the CP violating effects may be present in this limit, or even in the limit where $m_d=m_s$. 

CP-even WBIs can also be very useful in the study of the flavour structure of models with VLQs. In the SM, it has been shown that the knowledge of four specific CP-even WBIs allows one to reconstruct the full $V_\text{CKM}$ matrix with only a two-fold ambiguity in the sign of CP violation. In the specific case of an extension with a heavy up-type VLQ, the quark mixing matrix appearing in the weak charged currents is a $4\times 3$ matrix. We have identified the nine CP-even WBIs which determine this quark mixing matrix.

We have pointed out that in the presence of a VLQ, the lowest order CP-odd WBI is of mass order  $M^8$, much lower than the order of the CP-odd WBI in the SM (which is of order $M^{12}$). As a result, the strength of CP violation relevant for baryogenesis is much larger in models with VLQs than the corresponding one in the SM.
\clearpage	
	%%%%%%%%%%%%%%%%%%%%%%%%%%%%%%%%%%%%%%%%%%%%%%
	\section*{Appendix}
	\appendix

\section{Constructing WBIs for a general extension with VLQ isosinglets}
\label{WBIgen}

Although in this work we focused specifically on extensions with one up-type VLQ isosinglet, it is straightforward to understand that all the results presented are applicable to extensions with one down-type VLQ isosinglet just by interchanging the up and down sectors.

However, the formalism we presented in section \ref{sec:WBI} is also applicable to a general extension with $n_u$ up-type and $n_d$ down-type VLQ isosinglets. In this appendix we briefly demonstrate that.

In the general case, WBTs are defined in an analogous way to before. They correspond to arbitrary $3\times 3$ unitary transformations of the standard left-handed quark fields $u_L$ and $d_L$, to $n\times n$ (with $n=n_u$ or $n=n_d$, depending on the specific sector) unitary transformations of the left-handed VLQ fields $U_L$ or $D_L$, as well as $(3+n)\times (3+n)$ unitary transformations of the right-handed VLQ fields $u_R$ or $d_R$.

The matrices relating the flavor eigenstates with the mass eigenstates $u'$ and $d'$, are the $(3+n)\times(3\times n)$ unitary matrices defined as
\begin{equation}
 \mathcal{V} = \begin{pmatrix}
A \\
B
\end{pmatrix},
\label{AB_2}
\end{equation}
for each sector, where $A$ is $3\times (3\times n)$ and $B$ is $n\times (3\times n)$. The mass matrices $\mathcal{M}_u$ and $\mathcal{M}_d$ are diagonalized as
\begin{equation}
    \mathcal{V}^\dagger\mathcal{M}\mathcal{W}=\mathcal{D}
    \label{diagon_2}
\end{equation}
with $\mathcal{D}$ being the diagonal matrix containing the quark masses of a given sector and $\mathcal{W}$ having the same size as $\mathcal{V}$. The unitarity of each $\mathcal{V}$ means that the relations in Eq.(\ref{eq:AAdagger}) are still true for each sector individually.

Moreover, for each sector, the $3+n$ quark masses are determined by a set of $3+n$ invariants constructed from $\mathcal{H}_{u,d}=\mathcal{M}_{u,d}\mathcal{M}^\dagger_{u,d}$. These can be defined as
\begin{equation}
    \chi_\sigma\left(\mathcal{H}\right)=\frac{1}{\sigma!}\frac{\partial^\sigma}{\partial x^\sigma}\det\left(\mathcal{H}+x\mathbb{1}_{n\times n} \right)\Big|_{x=0},
\end{equation}
where $\sigma$ runs from $0$ to $2+n$, with $\chi_0\left(\mathcal{H}\right)=\det\left(\mathcal{H}\right)$ and $\chi_{2+n}\left(\mathcal{H}\right)=\tr\left(\mathcal{H}\right)$. 

The Lagrangian describing the charged currents is now given by 
\begin{equation} \label{eq:chargedcurrent_2}
    \mathcal{L}_W = \frac{g}{\sqrt{2}} W_\mu^+ \overline{u^\prime}_{L} A_u^\dagger A_d  \gamma^\mu d^\prime_L + \mathrm{h.c.},
\end{equation}
so that now we define the $(3+n_u)\times(3+n_d)$ mixing matrix as $V_\text{CKM}\equiv A^\dagger_u A_d$.

The neutral current Lagrangian is given by
\begin{equation} \label{eq:neutralcurrent_2}
    \mathcal{L}_Z = \frac{g}{c_W} Z_\mu \left[ \frac{1}{2} \left(\overline{u^\prime}_{L}\  A^\dagger_u A_u\  \gamma^\mu u^\prime_L - \overline{d^\prime}_{L}A^\dagger_d A_d \gamma^\mu d^\prime_L \right) - s_W^2 \left(\frac{2}{3} \overline{u^\prime} \gamma^\mu u^\prime - \frac{1}{3} \overline{d^\prime} \gamma^\mu d^\prime \right) \right],
\end{equation}
so that now there exist FCNCs on both sectors, which are controlled by the matrices
\begin{equation}
    \begin{split}
            F^u=& \ A^\dagger_u A_u = (A^\dagger_u A_d) (A^\dagger_d A_u)=V_\text{CKM}V^\dagger_\text{CKM},\\
            F^d=& \ A^\dagger_d A_d = (A^\dagger_u A_d) (A^\dagger_d A_u)=V^\dagger_\text{CKM}V_\text{CKM},\\
    \end{split}
\end{equation}
where we used the relation $AA^\dagger=\mathbb{1}_{3\times 3}$ coming from the unitarity of $\mathcal{V}$.

Now, to construct WBIs that relate to the mixing, as done before, it is useful to write for both sectors as (using Eq.(\ref{diagon_2}))
\begin{equation}
    \mathcal{M}=\begin{pmatrix}
    m\\
    M\\
    \end{pmatrix}=\begin{pmatrix}
    A\\
    B\\
    \end{pmatrix}\ \mathcal{D}\ \mathcal{W}^\dagger,
\end{equation}
From this we can define various hermitian matrices which transform as in Eq.(\ref{uuu}) under WBTs. Keeping the analogy with the one VLQ case, we can construct the set of matrices
\begin{equation}
    H^{(r)}\equiv m (m^\dagger m +M^\dagger M)^{r-1}m^\dagger= A \mathcal{D}^{2r} A^\dagger,
\end{equation}
which then can be used, just as in Eq.(\ref{eq:WBI}), to determine the moduli of all the entries of $V_\text{CKM}$ from
\begin{equation}
    \begin{split}
        \tr(H^{(r)}_u H^{(s)}_d)=& \ \tr\left(A_u \mathcal{D}^{2r}_u A^\dagger_u A_d \mathcal{D}^{2s}_d A^\dagger_d \right)=\tr\left(V^\dagger_\text{CKM}\mathcal{D}^{2r}_u V_\text{CKM}\mathcal{D}^{2s}_u\right) \\
         = & \ \sum^{3+n_u}_{\alpha=1}\sum^{3+n_d}_{\beta=1}\left(m^2_{u_\alpha}\right)^r\left(m^2_{d_\beta}\right)^s |V_{\alpha \beta}|^2.
    \end{split}
\end{equation}

The CP-odd WBI of dimension $M^8$ in Eq.(\ref{CP-odd}) can now be written as
\begin{equation}
   \begin{split}
    \tr\left([h_u,h_d]H^{(2)}_u\right)= & \ 2i \ \text{Im}\left[\tr\left(A_u \mathcal{D}^2_u A^\dagger_u A_d \mathcal{D}^2_d A^\dagger_d A_u \mathcal{D}^4_u  A^\dagger_u \right)\right]\\
     = & \ 2i \ \text{Im} \left[\tr\left( \mathcal{D}^2_u \ V_\text{CKM} \mathcal{D}^2_d \ V^\dagger_\text{CKM} \mathcal{D}^4_u F^u \right)\right]\\
     = & \ 2i \sum^{3+n_u}_{\alpha,\rho=1} \sum^{3+n_d}_{\beta=1} m^2_{u_\alpha}m^{2}_{d_\beta}m^4_{u_\rho}\text{Im}\left(F^u_{\rho \alpha}V^*_{\rho \beta}V_{\alpha \beta}\right),
    \end{split}
\end{equation}
It is important to note that now, in the general case, there is another distinct WBI of the same dimension. This can simply be obtained by interchanging the sector indices, \text{i.e.} $u\leftrightarrow d$, leading to
\begin{equation}
   \begin{split}
    \tr\left([h_d,h_u]H^{(2)}_d\right)= & \ 2i \ \text{Im} \left[\tr\left( \mathcal{D}^2_d \ V^\dagger_\text{CKM} \mathcal{D}^2_u \ V_\text{CKM} \mathcal{D}^4_d F^d \right)\right]\\
     = & \ 2i \sum^{3+n_u}_{\alpha=1} \sum^{3+n_d}_{\beta,\rho=1} m^2_{u_\alpha}m^{2}_{d_\beta}m^4_{d_\rho}\text{Im}\left(F^d_{\rho \beta}V_{\alpha \rho}V^*_{\alpha \beta}\right).
    \end{split}
\end{equation}

	\section{Simplified minimal set of Invariants for one up-VLQ} \label{apa}
	In this Appendix, we consider a WB and a simplified minimal set of invariants. Our aim is to exemplify, in an alternative but comprehensive way, how this minimal set can be used to determine all physical parameters.
	
Consider the following hermitian matrix
\begin{equation}
\overline{h}_{u}=m_{u}M_{u}^{\dagger }(M_{u}M_{u}^{\dagger
})^{-1}M_{u}m_{u}^{\dagger },
\label{huu}
\end{equation}	
as well as the combination
\begin{equation}
h_\text{eff}=h_{u}-\overline{h}_{u},
\label{huu1}
\end{equation}
which is the effective (squared) mass matrix that gives (in most cases) a very good approximation of the three light up-quarks masses and mixings. As can be checked, under WBTs, the Hermitian matrices in Eqs. (\ref{huu},\ref{huu1}) transform as the matrices $h_d=m_dm^\dagger_d$ and $h_u=m_um^\dagger_u$ defined in Eq.(\ref{eq:herm_matr}), \textit{i.e.}
$h \rightarrow V^\dagger\  h\  V$.

We use this minimal set of hermitian matrices to construct simple WB invariants. 
To do this, consider the WB where the up and down quark mass matrices are given by 
\begin{equation}
\begin{array}{lll}
\mathcal{M}_{u}=\left( 
\begin{array}{llll}
m_{1} & 0 & 0 & r_{1} \\ 
0& m_{2} & 0 & r_{2} \\ 
0&0  & m_{3} & r_{3} \\ 
0& 0 & 0 & m_{4}
\end{array}
\right)  & , & m_{d}=U D_{d},
\end{array}
\label{special_WB}
\end{equation}
where all entries of $\mathcal{M}_{u}$ are real. This form is always possible with adequate unitary transformations of the quark fields. 

The down-sector mass matrix contains the diagonal matrix $D_d$ with the three down quark masses and a $3\times 3$ unitary matrix $U$, which one can write as
\begin{equation}
    U= K_{\alpha'\beta'}O_{23}K_{\delta'}O_{13} O_{12}
    \label{Uhd}
    \end{equation}
where $K_{\alpha'\beta'}=\text{diag}(1, e^{i\alpha'}, e^{i\beta'})$, $K_{\delta'}=\text{diag}(1, 1, e^{i\delta'})$ and the $O_{ij}$ are basic real orthogonal transformations in the $ij-$plane. In this WB, the down sector has a total of 9 physical parameters: 3 mixing angles, 3 quark masses and 3 being CP violating phases. The unitary matrix $U$ is, for all physical relevant cases, very near to the $3\times 3$ part of the CKM mixing matrix.

With respect to the up-quark mass matrix (and again for the physical relevant cases), the $m_1, m_2,m_3$ are very good approximations of the lightest quark masses $m_u,m_c,m_t$.  

In this WB, one obtains very simple forms for the Hermitian matrices in Eqs. \eqref{huu}, \eqref{huu1}, \eqref{special_WB},
\begin{equation}
    h_d=U \ \text{diag}(m^2_d,m^2_s,m^2_b)\ U^\dagger.
    \label{hd1}
\end{equation}
and
\begin{equation}
    \begin{split}
    h_\text{eff}=& \ \text{diag}(m_{1}^{2},m_{2}^{2},m_{3}^{2}),\\
    \Bar{h}_{u}=& \ \text{diag}(r_{1},r_{2},r_{3})\cdot
\begin{pmatrix}
1 & 1 & 1 \\ 
1 & 1 & 1 \\ 
1 & 1 & 1
\end{pmatrix}\cdot \text{diag}(r_{1},r_{2},r_{3}).\\ 
    \end{split}
\end{equation}
Thus, $h_u=h_\text{eff}+\Bar{h}_u$ has also a very manageable form.

Alternatively, and in order to provide a more simplified description, we may switch from the parametrization of $h_d$ in Eq.(\ref{hd1}) to
\begin{equation}
    h_d=\begin{pmatrix}
    1 & 0 & 0 \\
    0 & e^{i\alpha} & 0 \\
    0 & 0 & e^{i\beta}\\ 
    \end{pmatrix}\begin{pmatrix}
    p & a & b e^{i\delta}\\
    a & q & c \\
    b e^{-i\delta} & c & r\\ 
    \end{pmatrix}\begin{pmatrix}
    1 & 0 & 0 \\
    0 & e^{-i\alpha} & 0 \\
    0 & 0 & e^{-i\beta}\\ 
    \end{pmatrix},
\end{equation}
where $a,b,c,p,q,r$ are all real parameters and $\alpha,\beta,\delta\in [0,2\pi]$. All these new parameters are evident functions of the nine physical parameters in Eq.(\ref{hd1}). Notice that we have the same number of real parameters and complex phases as before.

One may now relate all parameters in the mass matrices with WBIs constructed from $h_u$, $\overline{h}_{u}$, $h_\text{eff}$ and $h_d$. For instance, the quantities  $m_1$, $m_2$, $m_3$ can be obtained by using $\text{tr}(h_\text{eff})$, $\chi(h_\text{eff})$ and $\text{det}(h_\text{eff})$, while $m_4$ is obtained from $m^2_4=\text{tr}(\mathcal{H}_{u})-\text{tr}(h_{u})$, where $\mathcal{H}_{u}=\mathcal{M}_{u}\mathcal{M}_{u}^\dagger$.

Now, using this knowledge, one then proceeds to determine $r_1$, $r_2$, $r_3$ from
\begin{equation}
\begin{array}{lll}
\text{tr}(\overline{h}%
_{u}h_\text{eff})=m_{1}^{2}r_{1}^{2}+m_{2}^{2}r_{2}^{2}+m_{3}^{2}r_{3}^{2}, \\ 
\text{tr}(\overline{h}%
_{u}h_\text{eff}^{2})=m_{1}^{4}r_{1}^{2}+m_{2}^{4}r_{2}^{2}+m_{3}^{4}r_{3}^{2}, \\ 
\text{tr}(\overline{h}%
_{u}h_\text{eff}^{3})=m_{1}^{6}r_{1}^{2}+m_{2}^{6}r_{2}^{2}+m_{3}^{6}r_{3}^{2},
\end{array}
\end{equation}
or from three other invariants of
$\mathcal{H}_{u}$: $\tr(\mathcal{H}_{u})$, $\chi_1(\mathcal{H}_{u})$ and $\chi_2(\mathcal{H}_{u})$.

Subsequently, the parameters $p,q$ and $r$ can be determined from 
\begin{equation}
    \begin{split}
    \text{tr}(h_d)& =p+q+r,\\
    \text{tr}(h_d h_\text{eff})&  =m^2_1 p + m^2_2 q + m^2_3 r,\\
    \text{tr}(h_d h^2_\text{eff})&  =m^4_1 p + m^4_2 q + m^4_3 r,\\
    \end{split}
\end{equation}
and after that $a,b$ and $c$ are obtained with
\begin{equation}
    \begin{split}
    \text{tr}(h^2_d) = & \ p^2+q^2+r^2+2(a^2+b^2+c^2),\\
    \text{tr}(h_d h_\text{eff})  = & \ m^2_1 p^2 + m^2_2 q^2 + m^2_3 r^2\\
    & +(m^2_1+m^2_2)a^2+(m^2_2+m^2_3)b^2+(m^2_1+m^2_3)c^2,\\
    \text{tr}(h_d h^2_\text{eff}) = & \ m^4_1 p^2 + m^4_2 q^2 + m^4_3 r^2 \\
    & +(m^4_1+m^4_2)a^2+(m^4_2+m^4_3)b^2+(m^4_1+m^4_3)c^2.\\
    \end{split}
\end{equation}

Moreover, the complex phase $\delta$ can be extracted from
\begin{equation}
    \text{det}(h_d)=p q r - a^2 r - b^2 q -c^2 p + 2 a b c \cos\delta,
\end{equation}
whereas the remaining phases $\alpha$ and $\beta$ are determined from
\begin{equation}
    \begin{split}\label{eq:SWB_CP_odd}
    \frac{1}{2i}\text{tr}\left([h_d,\Bar{h}_u]h_\text{eff}\right) =& \ a (m^2_2-m^2_1)r_1r_2\sin\alpha +b (m^2_1-m^2_3)r_1r_3\sin(\beta-\delta) \\
    & +c(m^2_3-m^2_2)r_2r_3\sin(\alpha-\beta),\\
     \frac{1}{2i}\text{tr}\left([h_d,\Bar{h}_u]h^2_\text{eff}\right) =& \ a (m^4_2-m^4_1)r_1r_2\sin\alpha +b (m^4_1-m^4_3)r_1r_3\sin(\beta-\delta) \\
    & +c(m^4_3-m^4_2)r_2r_3\sin(\alpha-\beta).\\
    \end{split}
\end{equation}
Note that although $\text{tr}([h_d,\overline{h}_u]h_\text{eff})$ is a CP-odd WBI of dimension 6, this does not contradict the discussion of section \ref{sec:CPV}. In reality, using the notation of section \ref{sec:constr_invs} we have $\Bar{h}_u\equiv\left(M_u M_u^\dagger\right)^{-1}(H^{(2)}_u-h^2_u)$, meaning that
\begin{equation}
    \text{tr}([h_d,\overline{h}_u]h_\text{eff})=(M_uM^\dagger_u)^{-1}\text{tr}([h_u,h_d]H^\text{(2)}_u),
\end{equation}
so that the WBI in (\ref{eq:SWB_CP_odd}) has only such a low dimension due to a scaling factor $(M_uM^\dagger_u)^{-1}\sim 1/m^{2}_4$. It is not, therefore, a new WBI of lower dimension.

Thus, we have obtained all physical parameters of our model in terms of a minimal set WB invariants.

	\section*{Acknowledgments}
This work was partially supported by Fundação para a Ciência e a Tecnologia (FCT, Portugal) through the projects CFTP-FCT Unit 777 (UIDB/00777/2020 and UIDP/00777/2020), PTDC/FIS-PAR/29436/2017, CERN/FIS-PAR/0008/2019 and CERN/FIS-PAR/0002/2021, which are partially funded through POCTI (FEDER), COMPETE, QREN and EU.
	%%%%%%%%%%%%%%%%%%%%%%%%%%%%%%%%%%%%%%%%%%%%%%
	
	%%%%%%%%%%%%%%%%%%%%%%%
	
	\providecommand{\noopsort}[1]{}\providecommand{\singleletter}[1]{#1}%
	
	\providecommand{\href}[2]{#2}\begingroup\raggedright\endgroup
		
	\end{document}